\documentclass[sigchi]{acmart}
\AtBeginDocument{%
  \providecommand\BibTeX{{%
    \normalfont B\kern-0.5em{\scshape i\kern-0.25em b}\kern-0.8em\TeX}}}

\setcopyright{acmcopyright}
\copyrightyear{2024}
\acmYear{2024}
\acmDOI{}

\acmConference[]{}
\acmBooktitle{}
\acmPrice{15.00}
\acmISBN{978-1-4503-XXXX-X/18/06}

\usepackage{booktabs}
\usepackage{dcolumn}
\usepackage{tabularx}
\usepackage{framed}
\usepackage{graphicx}
\usepackage{tabularx}
\usepackage{graphicx}
\usepackage{url}
\usepackage{tabulary}
\usepackage{multirow}
\usepackage{enumitem}
\usepackage{hyperref}
\usepackage{amsmath}
\usepackage{lscape}
\usepackage[normalem]{ulem}
\useunder{\uline}{\ul}{}
\usepackage{longtable}
\usepackage{lscape}
\usepackage{natbib}

\usepackage{xcolor}
\usepackage{color, colortbl}

\usepackage{siunitx}
\usepackage{microtype}
\usepackage{etoolbox}
\usepackage{makecell}
\usepackage{subcaption}

\usepackage{multicol}
\usepackage{threeparttablex}
\usepackage{siunitx}
\usepackage{array}
\newcolumntype{P}[1]{>{\centering\arraybackslash}p{#1}}
\usepackage{multirow}
\usepackage{caption} 
\usepackage[tableposition=above]{caption}
\usepackage{adjustbox}
\usepackage{amsmath}
\usepackage[font=normalsize,labelfont={bf}]{caption}
\begin{document}



\title[Exploring Plural Perspectives in Self-Tracking Technologies]{Exploring Plural Perspectives in Self-Tracking Technologies: Trust and Reflection in Self Tracking Practices}

\author{Sujay Shalawadi}
\orcid{0000-0003-3937-5427}
\affiliation{%
  \institution{Aalborg University}
    \city{Aalborg}
  \country{Denmark}}
\email{sujaybs@cs.aau.dk}

\author{Rosa van Koningsbruggen}
\orcid{0000-0002-1652-2888}
\affiliation{%
  \institution{Bauhaus-Universit{\"a}t Weimar}
    \city{Weimar}
  \country{Germany}}
\email{rosa.donna.van.koningsbruggen@uni-weimar.de}

\author{Rikke Hagensby Jensen}
\orcid{0000-0002-0009-8840}
\affiliation{%
  \institution{Aarhus University}
    \city{Aarhus}
  \country{Denmark}}
\email{rhj@cc.au.dk}

\renewcommand{\shortauthors}{Shalawadi et al.}

\begin{abstract} 

Contemporary self-tracking technologies (STTs), such as smartwatches and smartphone apps, allow people to become self-aware through the datafication of their everyday lives. However, concerns are emerging over the global north/Western portrayal of the self in the envisionment of STTs. Given the call to diversify participant samples in HCI knowledge building, we see it timely in understanding the influence of ubiquitous STTs in global south societies. We conduct a between-group analysis of 156 and 121 participants from Global North and South through two iterative surveys, respectively. We uncover significant differences in perceived trust with their STTs and reflection practices between the groups. We provide an empirical understanding on advocating for inclusive design strategies that recognize diverse interpretations of STTs and highlight the need to prioritize local values and flexibility in tracking to foster deeper reflection across cultures. Lastly, we discuss our findings in relation to the existing literature and highlight design recommendations for future research.

\end{abstract}



\maketitle

\section{Introduction}

Contemporary self-tracking technologies (STTs), such as smartwatches and smartphone apps, have made monitoring personal experiences in daily life more convenient and frequent than ever before. These tools not only allow for easy tracking of various data metrics like steps, calorie intake, and heart rate but also enable users to analyze and interpret their data for increased self-awareness. This trend is commonly known as the quantified self (QS) movement~\cite{wolf2016quantified}. As a result, an increasing number of individuals worldwide are embracing the QS movement, using data to gain insights into individual health improvement strategies. However, a conflicting discourse questions the notion of the ``self'' in QS, portraying it as imperfect and driven by needs that are perceived to exist separately from body and mind~\cite{Hong2020, STTPractices, lupton2016quantified}.

Concerns arise over the portrayal of the self in most STTs, which predominantly treat individuals as subjects to be tracked and quantified, with data used for indexing and economic gain~\cite{neff2016self,zuboff2019age}. As developers strive to make STTs continuous, pervasive, and affordable globally, particularly in global south societies. The local, cultural, and intellectual traditions of these societies may not align with the STTs' intended impact \cite{Postcolonial, Merrit_PCLanguage_2011, Escobar2018, STTPractices}. Since STTs can shape and change local cultures and practices, it is crucial to recognize that the QS movement, promoted by STTs, is often seen as a Western concept due to its origin and prevalence in societies of the Global North~\cite{cetkovic2016refracted}. 

Given the profound impact that STTs can have on mediating our lives, previous research in the field of human-computer interaction (HCI) has diligently sought to understand the prevailing attitudes and practices concerning STTs to ensure their meaningful design ~\cite{PILit}. However, a gap persists in the literature, as many studies overlook the crucial consideration of the geographical backgrounds of participant samples in their investigations ~\cite{Idontneedagoal,PILit}. In response to Linxen et al.~\cite{CHIisWEIRD}'s call to diversify participant samples and increase the inclusion of global south individuals in HCI research, our study investigates the influence of STTs on self-trackers from geographically diverse backgrounds.

As Self-Tracking Technologies (STTs) continue to proliferate in the consumer market, they are increasingly embedded within diverse geographical contexts, each characterized by distinct daily rhythms and cultural practices~\cite{Idontneedagoal,CultureHome}. This expansion presents a unique opportunity for researchers to enhance the inclusivity and cultural sensitivity of STTs, ensuring they better align with the varied values of global users. Recognizing this potential, we focused our study on the crucial elements of trust and reflection in STTs, as these factors profoundly influence how users interact with and derive value from these technologies. Trust serves as the foundation that enables meaningful engagement with the technology~\cite{leuenberger2024track}, while reflection is the process through which users gain personal insights and make informed decisions based on the data collected~\cite{Balance}. Together, trust and reflection create a reinforcing cycle: trust fosters deeper reflection, and positive reflective experiences, in turn, strengthen trust.

To investigate these dynamics, we conducted two iterative surveys with participants from both the Global South (GS) and Global North (GN) regions. The initial survey, involving 156 participants (78 from GS and 78 from GN), aimed to assess perceived trust in STTs. This was followed by a second survey with 121 participants (61 from GS and 60 from GN), designed to explore how these users engage in reflection practices and their relationship with STTs. These surveys provide valuable insights into how trust and reflection function across different cultural contexts, offering guidance for designing more inclusive and effective STTs. Based on the analysis compiled from the two surveys, we offer the following contributions. 

\begin{enumerate}
    \item \textit{Key findings:} We observe significant disparities were found in the trust placed in STTs, with GS participants showing notably lower levels of trust compared to GN participants. This lower trust influences their cautious and selective usage patterns, leading to a preference for combining digital and analog tracking methods. GS participants engage in deeper, more contextually grounded reflection practices. Unlike GN participants who may rely solely on STT feedback, GS participants often critically assess how their mood, environment, and daily experiences impact their data, leading to more nuanced insights.
    \item \textit{Implications:} The study highlights the need for inclusive design strategies that address the diverse interpretations and needs of STT users from different cultural and geographical backgrounds. Our findings highlight how participants frequently form quasi-relationships with their STTs, revealing mismatches between preferences and the functionalities of these technologies, notably among our GS participants. 
    \item \textit{Opportunities:} We note a substantial age disparity among participants, with younger GS individuals possibly encountering expert-driven STTs prematurely, highlighting the importance for designers to assess the effects of such practices. Despite GS participants' inclination towards stationary reflection locations like their homes, they still expressed a preference for mobile tracking devices. We recommend leveraging insights from extensive HCI research to prioritize local values and encourage tracking flexibility to foster more profound reflection.
\end{enumerate}

\section{Related Work} \label{RW}

To guide our investigation into opportunities for better understanding inclusivity and sensitivity to diverse values, we draw inspiration from three HCI paradigms: third-wave HCI, contemporary technological philosophies, and trust and reflection in inclusive STT design.

\subsection{Contemporary STTs in HCI}
 Contemporary STTs (such as smartphone applications) and wearables (such as smart watches) extend human bodies and are typically present with people at any time of the day, resulting in frequent tracking of activities and converting experiences into data to, for instance, make people aware of their well-being~\cite{zuboff2019age,neff2016self}. The central actor in universal (one-size-fits-all) self-tracking cultures, such as the QS movement, is data rather than the people who provide it~\cite{KnowThyself1,KnowThyself2,KnowThyself3}. This pattern is best encapsulated in the popular goal-centric model within the QS philosophy, where an individual is required to achieve the goals that must be met~\cite{STTPractices}. Those models are typically designed to advocate a one-size-fits-all approach with the preconceived idea that tracking data can only be used to improve `the quantified-self', and any failure to reach the tracker's objectives is considered the `fault' of the user~\cite{crawford2015our}. By removing the user's role of subjective interpretation, this idea further desensitizes subjective sensations hidden behind numbers, such as the idea that a step tracked whilst feeling happy has the same value as a step taken whilst feeling miserable ~\cite{Fitter}. Further, because STTs are well-equipped to monitor numerous data details of daily human life, the QS philosophy will imply that these technologies are also well-equipped to give humans information to understand their position in the world~\cite{ReviewingReflection,RevisitingReflection,ReflectionFitness}.

Nonetheless, as digital technology becomes increasingly ubiquitous in many aspects of our daily lives, these technologies are no longer restricted to only certain types of users with specific usability `goals' in mind~\cite{GoodDesign}. With Bødkers' coining of the third-wave in HCI~\cite{ThirdWave}, we see more and more HCI studies that recognize the ubiquitous nature of digital technology as complex relationships between human beings, digital technologies, and social-cultural contexts~\cite{,10YearsLater}, and embrace these complexities to shape more meaningful experiences through `good' interaction design~\cite{GoodDesign}. Because these technologies are embedded in most parts of everyday life, bringing a third-wave HCI perspective to frame understandings of the human experience of contemporary STTs suggests that interactive technologies can no longer be considered a concern for merely global north countries.

\subsection{Contemporary Philosophies in Technology and STTs}
Third-wave HCI emphasizes how interactive technological artifacts are entangled within the complex web of cultural, social, economic, political, and institutional influences that shape user experiences~\cite{GoodDesign,SmartphoneDiscourse}. Exploring the implications of these intertwined factors within contemporary STTs requires employing various means, such as ideas, concepts, and tools~\cite{fallman2007persuade}. Contemporary philosophical perspectives on technology offer insights into navigating the plurality and complexity of technology in everyday life. The debate between utopian and dystopian ideologies surrounding technology has persisted since the twentieth century, paralleling its increasing pervasiveness~\cite{fallman2007persuade}. Our approach is influenced by Don Ihde's concept of the non-neutrality of technology, which is among contemporary theories of technology~\cite{GoodDesign,ihde2012technics}.

Ihde proposes a view of technology as non-neutral, challenging the notion that technologies are impartial tools serving human purposes, a perspective echoed in HCI literature~\cite{GoodDesign}. He highlights two key characteristics to illustrate this non-neutrality. Firstly, technologies often enhance certain features while diminishing others, altering our experiences of the world~\cite{ihde1990technology}. Secondly, they transform our experiences by simultaneously concealing certain aspects of reality~\cite{ihde1990technology}. Thus, technologies play a mediating role in our perceptual, experiential, and bodily encounters~\cite{ihde2012technics}. Ihde categorizes human-technology relations into three types: embodiment, hermeneutical, and alterity relations.

\textit{Embodiment} extends the user's bodily awareness, \textit{hermeneutics} transforms encounters with the world via technology, and \textit{alterity} involves the user's relation with the agency of technology. These relations illustrate the idea that no technology is static and can belong to multiple contexts~\cite{ihde1999technology}. Technologies, therefore, exhibit multistability, serving various purposes and embodying different meanings.

In self-tracking, Ihde's human-technology relations help understand the influence of STTs on users. For instance, step-counting features in smartphone apps or wearables embody non-neutrality: users record movements via steps (embodiment), interpret visualizations of their data (hermeneutics), and perceive success or failure in meeting goals (alterity). This dynamic interaction highlights how technology mediates users' experiences and influences their relation to the world, amplifying their pursuit of well-being.

\subsection{Trust and Reflection in Inclusive STT Design}

We integrate Ihde's concept of technology's non-neutrality into self-tracking technology (STT) design, aiming to explore diverse perspectives from global south societies often overlooked in HCI research~\cite{Escobar2018,CHIisWEIRD}. STTs, once limited to the West, are now widely accessible, raising important implications for underrepresented populations in HCI research~\cite{Idontneedagoal,PILit}. Understanding the socio-technical phenomena arising from the ubiquitous nature of contemporary STTs is crucial for informing inclusive design, especially for these underrepresented contexts~\cite{InsandOuts}. Inclusive design for STTs should consider diverse expressions of self-tracking practices worldwide, challenging the dominance of global north perspectives~\cite{Aino_Wellness_2008, Devito_2019_Spinwave, DesigningWellBeing}.

Trust is a pivotal element in the design and adoption of STTs, significantly influencing various aspects of user interaction, from initial engagement to long-term usage and reflective practices~\cite{Busch2021}. Users who trust STTs are more likely to use them regularly and follow their recommendations, which is essential to achieve the intended benefits of self-tracking~\cite{Busch2021}. Trust also facilitates deeper engagement with the reflective aspects of self-tracking, as users who trust the technology are more likely to rely on the data for meaningful self-reflection and personal insights~\cite{McKnight}. This deep engagement is crucial for understanding one's behavior patterns and making informed decisions~\cite{barcena2014safe,huckvale2015unaddressed}.

When designing STTs with a focus on building trust, it is also important to consider and engage with the existing discussions and complexities surrounding the concept of reflection~\cite{ReviewingReflection}. This includes understanding how different users reflect on their data, the benefits and potential drawbacks of reflection, and the cultural nuances that influence how reflection is perceived and practiced~\cite{Li2010, CurseQS, TMRM, Epstein, Rooksby2014, ReviewingReflection, RevisitingReflection, Balance}. Essentially, building trust in STTs is not only about making the technology reliable and secure but also about ensuring that it supports meaningful and culturally sensitive reflective practices.

Although trust and reflection are commonly considered beneficial to the individual, recent HCI research highlights their dark sides and cultural nuances~\cite{ReviewingReflection,TMRM}. Yet, little research explores how underrepresented participants perceive reflection. Our study aims to reduce this gap by capturing opinions on perceived trust and local reflection practices from STT users, who are typically overlooked participants in HCI research.

\section{Methodology Overview}

To explore the roles of trust and reflection in STTs across diverse cultural contexts, we conducted a two-phase study involving participants from the Global South (GS) and Global North (GN). The first survey assessed perceived trust in STTs among 156 participants (78 from each region). Building on these insights, a follow-up survey with 121 participants (61 from GS and 60 from GN) examined how trust influences reflection practices. This dual-survey approach provides a comprehensive understanding of how trust and reflection interact, guiding the effective and inclusive design of STTs.



\section{Survey 1: Perception of trust in Self-Tracking Technologies} \label{Trust}

The purpose of this survey is to compare how people in the GN and GS regions trust the various STTs they commonly use to become self-aware. Although access to universal STTs such as smartphone applications and wearable trackers that embody QS philosophy are easily accessible, there exists an unequal power relationship between people who use STTs and the developers/designers of these technologies, which leads to unjust and biased designs~\cite{Dourish_ColonialImpulse_2012,neff2016self,lupton2016quantified, costanzachock2020designjustice}. We acknowledge that there are several aspects of human experiences or attitudes that can be influenced by this unequal power relationship. In this survey, we focus on perceived trust as a starting point to critically assess the implication of an unequal power relationship between people and developers/designers of STTs. 

\subsection{Procedure, Participants and Analysis}
We opted to use the Human-Computer Trust Scale (HCTS), a pre-validated scale developed by Gulati et al.~\cite{gulati2019design}. This scale has demonstrated consistency in various contexts, including smart home intimacy~\cite{DesignFiction}, ethics of computer supervision/support/surveillance, reliability of data for student learning activities~\cite{NewSchool}, and human-robot interaction~\cite{PortugueseHRI}. The HCTS comprises five subscales: perceived risk, competence, benevolence, reciprocity, and trust. It consists of 12 questions (see Figure~\ref{fig:HCTS} in Appendix ~\ref{Premise}), focusing on the ethical considerations and morality of technology designers towards end-users. Participants rate each question on a 5-point Likert scale (1 - Strongly disagree, 5 - Strongly agree). To minimize order effects, the sequence of the 12 questions was randomized for each participant. An attention check question was also inserted between these 12 questions to increase the response reliability.

The survey began with an introduction to STTs and their role in maintaining daily well-being (see Appendix~\ref{Premise}), providing context for potential participants. In addition to the 12 questions, participants were asked one open question: `What is your recipe for maintaining a healthy well-being in day-to-day life?' and one clarification question to list all the various STTs (both digital and analog) they are currently using.

We used Prolific, an academic crowd-sourcing platform, to recruit participants. We required that our potential participants spoke fluent English and had at least a 95\% acceptance score from their previous participation on the platform. Demographic data collected for the study cannot be traced in any way to the identity of the participant. Participants were compensated according to the rates recommended by the online platform. For this survey, we compensated \pounds 1.20 per participant, where the average time spent on the survey was 4.5 minutes.

To support our between-group analysis, we collected responses (N~=~156) of 78 each from GN and GS. Among both groups, self-identified gender was equally balanced within both groups (Male~=~39, Female~=~39). The average age of the participants in GN was 37 years (SD~=~15), and in GS, it was 25 years (SD~=~5,1). GN consisted of participants from 4 countries: United Kingdom (68), Germany (3), Ireland (2), and the United States (5). GS consisted of participants from 3 countries: Mexico (45), India (3), and South Africa (30). We acknowledge that our interpretations of clustering the listed countries into the GN and GS are merely a `rough stand-in' for a very small representation of respondents. However, within HCI research, the countries that we included in GS are typically under-represented~\cite{CHIisWEIRD}. When examining countries by ratio, which assesses their impact on HCI research relative to their population sizes, we note the following rankings for the selected countries within the GS group: Mexico, South Africa, and India hold positions 68th, 54th, and 79th, respectively, among the 93 countries listed that contribute to HCI knowledge. In contrast, within the GN group, the USA, Ireland, Germany, and the UK rank significantly higher, holding the 12th, 14th, 18th, and 7th positions, respectively, among the same listed countries.

We evaluated the perceived trust scores for statistical significance between the participant groups. Additionally, the free-text response to the single open-ended question was manually coded by two authors, resulting in 53 codes and 5 categories. Following extensive discussions within the entire team, a consensus was reached in highlighting two overarching themes: (1) well-being practices and (2) self-tracking practices. Since every participant correctly answered the attention check question, we considered all responses reliable for our analysis.

\subsection{Findings}

The results of our survey 1 are presented through a between-group analysis of self-trackers from GS and GN. We begin by highlighting the two identified themes and then present the statistical analysis of perceived trust scores. 

\subsubsection{Well-Being Practices.}

Both groups prioritize their daily well-being by establishing routines and organizing tasks or activities to maintain a healthy lifestyle. However, perceptions of what constitutes a successful routine vary within each group, particularly concerning the timing of goal achievement in contexts such as fitness and dietary habits. GS participants expressed prioritization of tasks or activities that are emotionally meaningful, eventually fulfill them over longer periods of time, which may not focus on short-term success, particularly for that specific day, ``\textit{By working hard and doing things that bring me joy and fulfillment. Doing one task every day that makes me happy and remember what I fight for (sic)}'' (P39-GS). Although GN participants also expressed value in obtaining success in achieving specific tasks, they tended to focus on the means that might bring success for that day, rather than the feelings such success might bring about ``\textit{Eat 5 fruits and vegetables as part of a balanced diet, enjoy 2 cups of coffee and take 10k steps }'' (P73-GN).

Another trend observed in both groups for maintaining a healthy well-being involved abstaining from perceived `negative' behaviors associated with digital devices. Participants expressed a preference to invest their time in meaningful activities that provided them with the desired fulfillment, rather than engaging excessively with smartphones and social networks. However, participants from GS emphasized the significance of allocating more time to social interactions, as expressed by one participant: ``\textit{Staying positive and paying attention to other things than my smartphone only, things such as exercising and spending time with family}'' (P30-GS). In contrast, GN participants prioritized self-improvement, with one participant stating: ``\textit{Taking time off social media and my phone, and exercising regularly to get healthy}'' (P33-GN).

\subsubsection{Self-Tracking Practices.}

Three codes emerged - journaling, repurposing, and refocusing - to describe the self-tracking practices. These codes illustrate the overarching effort of our participants to balance tasks and activities in their daily lives, aiming to become better self-optimized versions of themselves. This phenomenon aligns with the conceptualization of self-optimization by Gimpel et al.~\cite{gimpel2013quantifying}, where individuals strive to compete against their quantified selves to achieve perceived improvements. Additionally, both groups demonstrate significant overlaps in their STTs, particularly in smartphone applications (GN: 23, GS: 27) and wearable devices (GN: 13, GS: 26), predominantly used for tracking daily task organization and fitness.
 
Journaling was primarily used by GN participants through analog methods like paper diaries to capture reflections and experiences. For example, a participant in the GN group mentioned using a paper diary to document instances like headaches resulting from continuous work without breaks: ``\textit{Paper diary to note down when I get headaches from working continuously with no breaks}'' (P25-GN). However, GS participants journaled primarily on digital mediums like smartphones and desktops, ``\textit{Just a simple checklist app and my agenda (paper notebook)}'' (P71-GS).

Re-purposing was predominantly observed among GS participants, who integrated two or more STTs into their daily routines to enhance their subjective tracking by augmenting the data to better reflect their individual experiences. For example, a participant in GS combined a smartphone application with a digital calendar to track and journal her menstrual cycle, as expressed by the quote: ``\textit{I use Notion for organizing and My Calendar to track my menstrual cycle}'' (P61-GS). Furthermore, GS participants repurposed their STTs to facilitate recording their thoughts and reflections orally through audio messages. One participant described using a WhatsApp chat as a convenient tool to jot down daily notes or routines, stating: ``\textit{I use a WhatsApp chat with myself, it's a quick access tool, on that I can write my daily notes or routines}'' (P74-GS).

Refocusing was prominently discussed concerning smartphone applications. For example, a participant in GS explained how they use an app to increase awareness and reduce time spent on distracting apps: ``\textit{Refocusing was a big deal for me. I'd always get sidetracked by apps on my phone. But now, I've got this focus plant app that tracks my screen time and reminds me when I'm spending too much time on Instagram}''(P45-GS). Another participant in GN also explained how their wearable technology intervenes when they loose track of time,  ``\textit{I'd often lose track of time on Instagram. But thanks to my Fitbit's screen time tracker, I've become more aware. Now, I'm more mindful of how I spend my time online}'' (P12-GN).

\subsubsection{Perceived Trust in STTs.} 

Despite varied lifestyles and routine choices, our participant groups shared common STTs like smartphone applications and wearable devices. In cases of similar tracking interests, both groups had mutually overlapping STTs. Using this overlap, we compared participants' perceived trust in their STTs.

\begin{figure}[h!]
    \centering
    \includegraphics[width=0.49\linewidth]{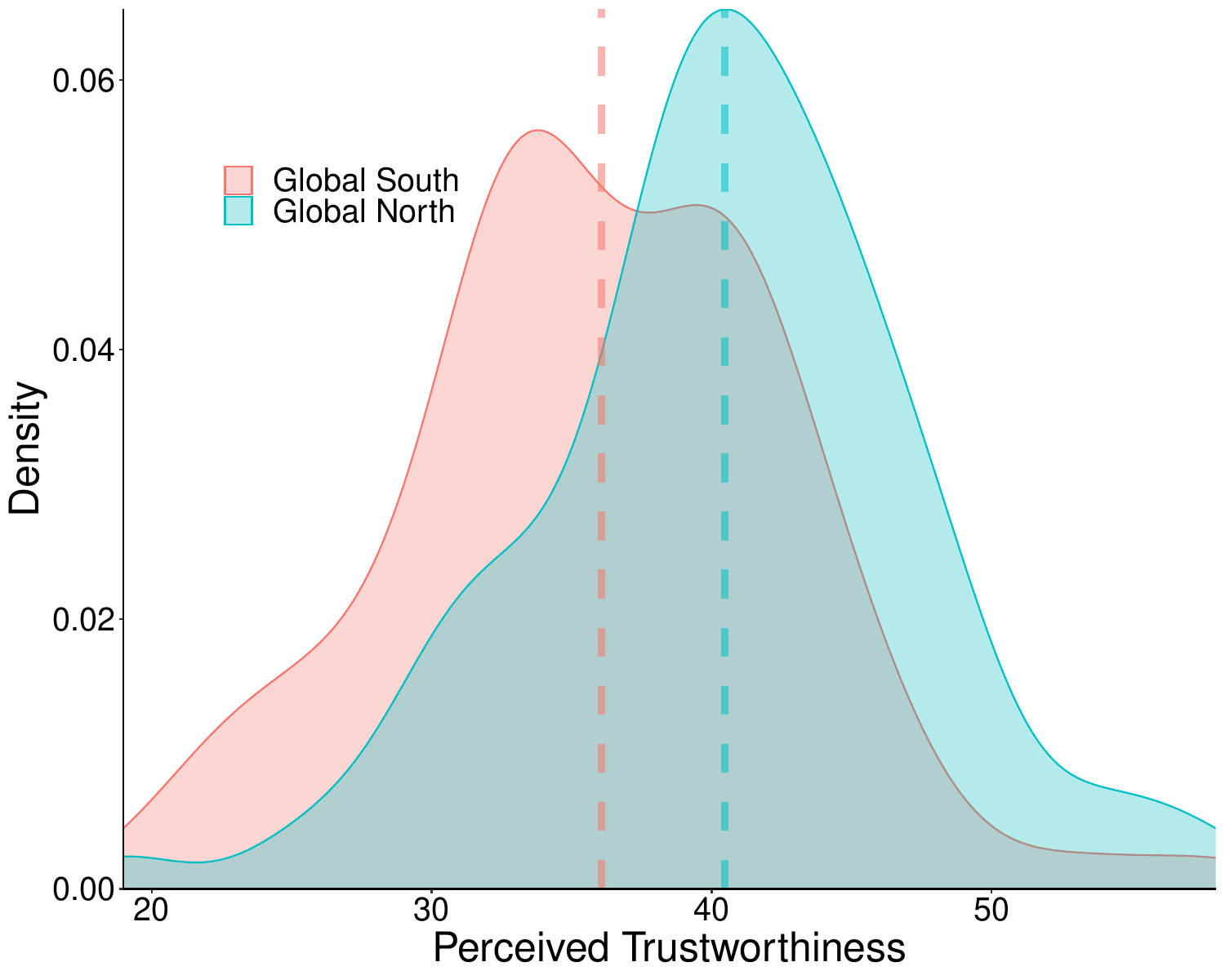}
    \caption{Density plots showing trust score distributions between the GS and GN. The vertical lines represent the median values (36 and 41) respectively for the GS and GN.}
    \label{fig:Distributions}
    \Description{Density plot distribution of the survey results, with vertical lines for median values. The distributions are overlapped to provide a visual of comparison between the two groups.}
\end{figure}

Perceived trust scores were computed using the methodology outlined by Gulati et al.~\cite{gulati2019design}. The results revealed that the GS group had lower overall trust scores in their STTs (M~=~36.06, SD~=~6.97) compared to the GN group (M~=~40.46, SD~=~6.9). The Likert scale questions and responses for both groups are detailed in Figure~\ref{fig:HCTS} in Appendix ~\ref{Premise}. 

A Kolmogorov-Smirnov normality test was conducted, resulting in p-values of 0.701 and 0.161 for the GS and GN groups, respectively, indicating a normal distribution. Density plots illustrating perceived trust scores for both groups are provided in Figure~\ref{fig:Distributions}.

Furthermore, the Levene test for the equality of variance yielded a p-value of 0.683, suggesting the equality of variance between the samples. A two-tailed t-test for independent samples (assuming equal variances) revealed a statistically significant difference between the GS and GN groups in perceived trust with their STTs (t(154)~=~-3.96, p < .001, 95\% confidence interval [-6.6, -2.2]). In other words, GS participants express significantly lower trust in their STTs compared to GN participants.

\section{Survey 2: Reflection Practices with Self-Tracking Technologies}

Our primary objective was to gain a better understanding of how perceived trust in STTs influences people's diverse and cultural reflection practices and to ascertain whether their use of STTs positively contributes to or impedes their overall well-being. Building on our related research, which has scrutinized the design of STTs, critiquing the notion of reflection as universally beneficial, and acknowledging its potential dark sides, such as the recall of negative memories \cite{InformingReflection,ReviewingReflection}, we sought to deepen our exploration of this discourse on reflection. To achieve this, we collected insights from self-trackers through our follow-up survey.

\subsection{Procedure and Participants}

\begin{figure}[h!]
    \centering
    \includegraphics[width=\linewidth]{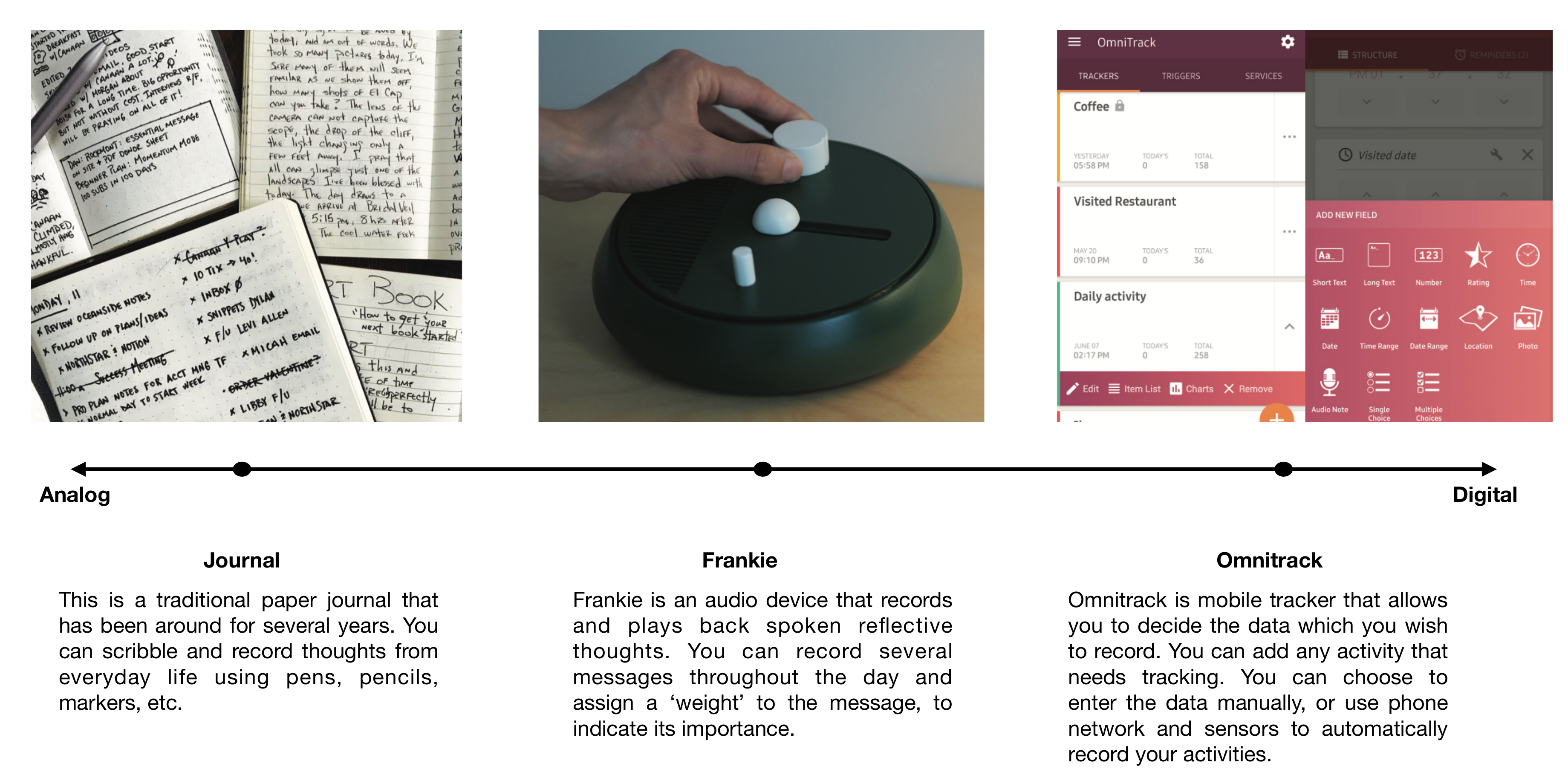}
    \caption{The three well-being technologies with explanations which were presented to the participants of SRTF. This visual was used to give participants a reference when thinking about different qualities of well-being technologies (e.g., flexibility).}
    \label{fig:Survey_Picture}
    \Description{3 pictures on a scale from analogue (left) to digital (right). Most left is a picture of handwritten journals, middle is Frankie --a tangible tracking prototype with 3 knobs, a hand touches one of the knobs--, an right is a screenshot of a tracking application.}
\end{figure}

We developed a survey inspired by Mols et al.~\cite{InformingReflection} consisting of 30 questions with an equal mix of quantitative and free-text questions (see Appendix~\ref{Reflection}). Quantitative questions comprised Likert scales, check boxes, and radio button responses, while qualitative questions had free-text responses. The premise of this survey was to provide participants with three examples of STTs that varied in the trade-off between \textit{flexibility} and \textit{convenience} in data tracking. Flexibility refers to the system's adaptability and customization options, while convenience relates to the ease and efficiency of data entry, management, and analysis processes. The examples were: 1) a paper diary, 2) Frankie: A tangible audio recording/playback artifact~\cite{Frankie} and 3) Omnitrack: A smartphone app that offers multiple ways to record data~\cite{OmniTrack}. In a broader sense, the three examples also varied on a spectrum of analog and digital mediums (see Figure~\ref{fig:Survey_Picture}). These examples are presented to encourage participants to reflect on their tracking habits and interaction preferences with their STTs.

\begin{itemize}
    \item A paper diary provides limited flexibility, relying on manual data entry without customizable parameters or automation. While it offers moderate convenience for traditional journaling enthusiasts, it demands manual effort for data input and lacks automated features for management and analysis.
    \item Frankie, allowing users to record spoken reflections, offers moderate flexibility compared to traditional written methods. It simplifies data capture through spoken recordings, enhancing convenience by reducing time and effort for entry. However, users may need to transcribe or review recordings manually for analysis, adding potential extra effort.
    \item Omnitrack offers high flexibility, customizable for tracking a wide range of data using sensors or user-reported information. With automated data capture and management features such as sensor-based tracking and data synchronization, it offers high convenience, enabling easy input, access, and analysis, thus minimizing manual effort and streamlining tracking processes.
\end{itemize}

\subsection{Participants and Analysis}

We extended invitations to participants from the previous survey. In total, we garnered responses from 121 individuals for this follow-up survey -- 61 from the GS group and 60 from the GN group. Despite the smaller sample size in the follow-up survey compared to the perceived trust survey, the reflection practices survey maintains a representation of the perceived trust sample. Although some participants dropped out and provided unreliable responses for analysis, the overall composition of participants in the reflection practices survey remains balanced across genders and reflective of the demographic characteristics observed in the perceived trust survey.

The average age of participants in the reduced sample for the GN group was 36 (SD~=~13), while for the GS group, it was 26 (SD~=~4.7). The GN group comprised participants from the United Kingdom (37), Germany (15), and the United States (3), while the GS group included participants from Mexico (31), South Africa (27), and India (3).

The free-text responses underwent thematic analysis, following a systematic process outlined by Braun and Clarke~\cite{TA}. Initially, all free-text responses were automatically compiled using atlas.ti (https://atlasti.com), with manual correction of phrases of uncertainty. Subsequently, we familiarized ourselves with the data through careful reading, marking noteworthy words and phrases and annotating them with short open codes. For instance, a quote from participant P3-GS, ``\textit{I start thinking about the things that I did}'', received the code `setting the scene'. After familiarizing ourselves with the data from both groups, we discussed our initial codes and iterated on them. Following a reflection period of one day, we refined the codes further, resulting in a revised set of 80 codes for the GN sample and 105 codes for the GS sample. Once the coding was completed, we began generating themes. During this phase, similar codes were grouped into clusters, forming candidate themes. This iterative process was repeated three times, culminating in the identification of three primary themes presented in the next section.

\subsection{Findings}

We present the three themes identified in this survey. The three themes are related to (1) emotion tracking and feedback asymmetry, (2) reflection strategies and settings for self-optimization, and (3) autonomy and convenience in plural data tracking.

The first theme contrasts how GN and GS participants view emotion tracking with their STTs, highlighting trust issues. GN participants express reservations due to concerns about privacy and the ambiguous nature of emotions, while GS participants value social support for specific goals like fitness. In the second theme, GN focuses on mental well-being, while GS prioritizes self-awareness and actionable insights. Both engage in well-being activities like music listening and meditation, but GS experiences spontaneous reflection and seeks distraction-free environments. Paradoxically, GS participants desire dedicated tracking devices like Frankie for uninterrupted reflection. In the third theme, GS participants appreciate Frankie's specificity and manual tracking despite additional effort, valuing autonomy related to accessibility and privacy. Conversely, GN participants value autonomy for self-expression and control over information, as reflected in their preference for Omnitrack. We supplemented our thematic analysis with quantitative data when needed.

\subsubsection{\textbf{Emotion Tracking and Feedback Asymmetry.}}

GN participants expressed a lack of interest in tracking emotions, stating that they are fuzzy, private, and cannot be accurately captured. However, the GS participants expressed that tracking emotions give you a better understanding of yourself and your well-being. GS participants stated that tracking emotions ``\textit{gives you the chance to comprehend what is happening inside you}''~(P27-GS) and that ``\textit{it is a good thing to always take out what's on your chest by writing it down}''~(P14-GS). On the contrary, the GN participants stated that it ``\textit{sounds awful to me}''~(P19-GN), and stressed their concerns about tracking emotions in the following two quotes: ``\textit{I don't think emotions can be accurately captured}''~(P4-GN) and ``\textit{Emotions are fuzzy and difficult to define. And they are private in a way that steps/activity is not}''~(P9-GN).

One potential explanation for this contrast might be the skepticism among GN participants regarding the ability of digital technologies to handle emotions. As one participant expressed, ``\textit{Traditional forms like [paper diary] journals better capture my emotions than technological devices}''~(P15-GN), while another stated, ``\textit{That seems far too personal to trust an app with}''~(P23-GN) in the context of tracking emotions. Additionally, participants voiced concerns that relying on apps could lead to detachment from their well-being, with one mentioning, ``\textit{I would feel detached from well-being as I can often spend time on my phone browsing online}''~(P5-GN).

GS participants reflected on how the tracking and goal-driven features of their STTs could boost motivation: ``\textit{It is cool to have goals that you can track and achieve on a daily or weekly basis it helps to have a motive or drive to keep going}''~(P2-GS), and clarify potential effects: ``\textit{My apps clearly specify me if I am working out this much, eating this much, I am going to improve that much}''~(P7-GS). This was supported by another participant who used feedback from their STTs to stay motivated and inspired: ``\textit{After I set a goal, I put myself under pressure to achieve it within the located time. When the app recognizes and compliments me on a goal I achieved, it makes me happy and inspired to do more}''~(P9-GS). GS participants also highlighted the social connections supported by their STTs as a core characteristic: ``\textit{Networking is an opportunity to create social connectedness, which means you have a great support system. The leaders are awesome, and they encourage personal growth and strive for goal achievement}''~(P29-GS).

The reluctance among GN participants to trust STTs with their emotional well-being contrasts with the dissatisfaction expressed by GS participants regarding the effectiveness of push notifications that were perceived as irrelevant and ineffective in the pursuit of wellness practices. This sentiment was evident for GS participants compared to GN participants, ``\textit{my tools felt like requirements instead of facilitators}''~(P5-GS), and pressure to achieve more or faster---as illustrated by the following quotes: ``\textit{I really like the weekly stats that it offers, but I felt anxious with my results and stress to make a change quicker}''~(P21-GS) and ``\textit{I would see peers posting their daily achievements and I would envy them}''~(P41-GS). A GS participant reflected on the number of notifications sent: ``\textit{I use a habit-building app, but it sometimes overwhelms me with the amount of notifications they send}''~(P1-GS). Other GS participants expressed annoyance that these notifications were generic, standardized, and impersonal: ``\textit{notifications from my mobile app which are irrelevant to what I do and how I want to use the app}''~(P12-GS), serving the purpose of someone else ``\textit{too many adverts online offering instant results for money}''~(P38-GS), and did not adapt to what they wanted to achieve as an individual ``\textit{recommendations that are general and not specific to me}''~(P4-GS).

Lastly, GN participants showed greater agency in their STTs because they helped mediate personalized fitness data in their wellness practices through specifically designed features: ``\textit{my Fitbit has a mindfulness section}''~(P12-GN). At the same time, they found their STTs trustworthy due to straightforward usability design aspects that ensured and `maximized' their well-being ``\textit{I don't allow any of the above to deter me; I take from the tools all and only what I need from them}''~(P54-GN). The increased agency displayed by GN participants in self-tracking, possibly due to factors such as improved technological access, familiarity with digital tools, and cultural norms emphasizing individual autonomy and personal health responsibility, likely influences their greater trust in these technologies compared to GS counterparts.

\subsubsection{\textbf{Reflection Strategies and Settings for Self-Optimization.}}
 
The approaches to reflection practices diverged noticeably between the two groups. GN participants observed that their reflection practices show a greater focus on self-care: ``\textit{feeling well and self-care, e.g., exercise frequently and eat food that nourishes the body}''~(P2-GN), whereas GS focused more on reflection as a means to improvement: ``\textit{It signifies enhancement in both mental and physical well-being. I contemplate my goals and aspirations, envisioning what I aim to accomplish for myself and my family.}''~(P31-GS). 

\begin{figure}[h!]
    \centering
    \includegraphics[width=\linewidth]{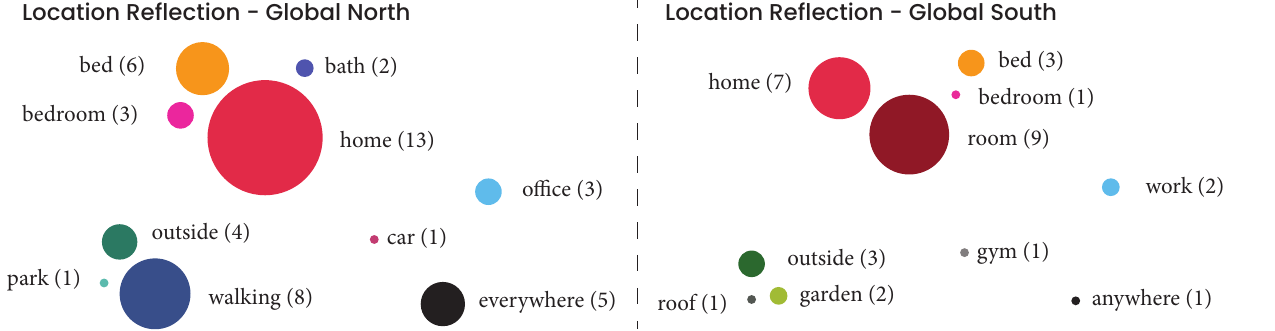}
    \caption{Overview of participants' locations of reflection. Each circle represents a location, which is written next to it. The size of the circle corresponds with how often that location was mentioned. The exact number can be read between the brackets, e.g., `home (13)'. This indicates that home was mentioned thirteen times.}
    \Description{A visualisation where the left side shows the location of GN and right of GS. The GN circles read: home (13), bed (6), bedroom (3), bath (2), office (3), car (1), everywhere (5), outside (4), walking (8), and park (1). GS circles read: room (9), bedroom (1), home (7), bed (3), outside (2), garden (2), roof (1), gym (1), anywhere (1), and work (2).}
    \label{fig:Location_Reflection}
\end{figure}

We observe activities that contribute to a reflective mindset are: listening to music, writing, meditation, breathing exercises, dialogue (alone or with others), and bodily activities, such as running, yoga, and walking. Among these activities, listening to music and walking were the most popular, stating that music brings back memories: ``\textit{When I listen to music, it often brings back memories or it speaks to my current life situation and that brings about reflection of my life}''~(P19-GS). The timing of reflection could be seen to be more spontaneous in the GS sample, occurring without any triggers: ``\textit{It just happens}''~(P14-GS). Another difference that could be constructed is \textit{`a moment of transcendence'}, which was present in GS, but not in GN. As described by GS participants, reflection is a state in which their ``\textit{mind just goes}''~(P1-GS), where they ``\textit{seem to fall into a sort of daydream}''~(P52-GS), and ``\textit{start drifting into my thoughts}''~(P2-GS). A visualization of the locations mentioned by the participants for their reflection can be seen in Figure \ref{fig:Location_Reflection}. We observed a paradox among the GS participants. They described their setting of reflection as a stationary moment, which happens mostly indoors in a private space while they are alone. In contrast to this setting, participants in this group also described the desire to have a mobile tracking device that allows easy data tracking anywhere and anytime. 

Most of the participants in both groups mentioned reflecting alone in a quiet place (e.g., home), where they feel comfortable. GS participants explicitly desired the absence of distractions: ``\textit{[I] look for a space where I can be free of distractions, in silence with myself}''~(P11-GS) and ``\textit{I usually turn off the wifi because the notifications usually distract me}''~(P27-GS). To achieve a moment without distractions, the GS participants highlighted the (potential) strengths of analog and dedicated devices [Frankie] for specific activities: ``\textit{I think it's important that it's a separate device from the cell phone to avoid distractions, plus give the gadget a different quality in your life}''~(P11-GS) and ``\textit{I like having devices that are specifically designed for one purpose}''~(P5-GS). 

\subsubsection{\textbf{Autonomy and Convenience in Plural Data Tracking.}}

Autonomy centered on the paper diary medium of tracking and was perceived differently between the groups. GN participants expressed autonomy more focused on freedom of expression and having the control to add desired information, ``\textit{A journal is the most basic and flexible of the options listed - it can be anything you want it to be and is analog and calm.}''~(P9-GN).  Although GS participants addressed the autonomy of a paper diary more in relation to their everyday context, ``\textit{you can take a journal anywhere, it doesn't need internet or batteries to work, you can come back to it whenever you want to}''~(P12-GS). GS participants also addressed autonomy in relation to their privacy compared to GN-participants, ``\textit{I value privacy and the liberty to decide when I would want to share my thoughts perhaps with a friend or close family or a professional help.}''~(P23-GS). Furthermore, the GS participants were supportive of the use of Frankie for its specificity compared to GN participants, ``\textit{I prefer using specialized devices, such as Frankie, that are designed for a specific purpose rather than relying on my phone for all tasks.}''~(P45-GS). 

Omnitrack was seen as a convenient option due to people's hectic daily schedules, ``\textit{I don’t always have the time to sit and write so a journal may not be used to it’s full potential. An application is quick to open and log activity whilst managing everyday life}''~(P17-GN). GN participants preferred transferring agency of tracking to their devices, ``\textit{I know I won't keep a journal up. It also records things automatically, which is again better because I don't remember to record things/have time to devote to this}''~(P9-GN). Although GS participants were aware of the additional work of manual tracking (e.g., through a paper diary), they expressed their satisfaction in undertaking such labor: ``\textit{I'm not interested in data, I wanna hear a story that makes me feel interested}''~(P1-GS).

GN participants expressed the convenience of Omnitrack's automatic nature of capturing data, but also of Omnitrack's freedom to choose any data type to record: ``\textit{the tracking application will record actions automatically as they occur. Little effort is needed, including adding media such as photos, screenshots, activities which I may find would serve}''~(P7-GN). However, it was apparent that GS participants favored Frankie over Omnitrack for tracking data: ``\textit{with the app, it automatically records. The recording [from Frankie] allows you to decide what to record and when}''~(P9-GS). Furthermore, the GS participants mentioned the advantage of the Frankie audio modality as it is more reflexive to their thoughts and offer higher flexibility than analogue writing: ``\textit{Talking to a recording device is very flexible because you can only think about it and say it, no need to spend an extra effort on writing down or keeping the record on an app.}''~(P24-GS). 

The desired ability to customize any kind of data and the perceived nature of Omnitrack to be universal and capable of tracking and offering new perspectives were widely seen in both groups: ``\textit{Sometimes its quite hard to know the why and how and technology can give a different perspective unlike the other 2 [compared with paper diary and Frankie]}''~(P2-GN). GN participants were more explicit than GS participants on Omnitrack to offer an overview of the tracked data and the automatic nature of tracking: ``\textit{I think this [Omnitrack] would be a more complete overview as some things would be automatically recorded}''~(P29-GN). The GS-participants considered Omnitrack more an obvious choice due to the `smartness' linked to the device: ``\textit{Omni track is a smart device unlike a journal and audio device}''~(P30-GS). GS participants also preferred the audio mode as an authentic or rather specific characteristic of Frankie compared to GN participants: ``\textit{I think I would choose Frankie because currently I consider that it is easier to communicate quality and quantity through voice, maybe we would have to sort this information and catalog it later, but in terms of storage, this option gives much more freedom than writing by hand or an app that tracks automatically.}''~(P11-GS). GN participants favored Omnitrack for audio tracking, which was integrated into the application along with other types of data.

\section{Discussion}

In this paper, we examined the differences in how individuals from the global north and south who engage in self-tracking, trust their self-tracking technologies and how these technologies facilitate or impede their reflective practices. We used two iterative surveys to collect insights to achieve an empirical understanding of self-tracking technologies and the notion of well-being and reflection practices. We illustrated that both groups differ significantly in their perceived trust in the STTs they use. Based on these insights, we emphasize that fostering meaningful well-being practices entails more than just automating data tracking with multi-purpose devices like smartphones and compelling reflection. We discuss different aspects of these complexities in the following sections. 

\subsection{Interplay of Trust and Reflection Practices}

Our findings suggest that the lower trust in STTs among GS participants not only affects their usage patterns but also influences the depth and nature of their reflection practices. The quantitative results from survey 1 indicate that GS participants report significantly lower trust in STTs compared to GN participants. This lower trust manifests in more hesitant and cautious usage patterns. GS participants might be less inclined to rely on STTs for critical health-related tasks, preferring to use them in a limited capacity or in conjunction with more traditional methods. For example, while GN participants might fully integrate STTs into their daily routines, GS participants might use STTs sporadically or only for specific functions they deem reliable.

Furthermore, qualitative data from survey 1 and 2 reveal that GS participants often repurpose STTs to fit their specific needs, indicating a selective and creative use of these technologies. Due to their lower trust, GS participants might combine STTs with other tools to improve their reliability. For instance, they might use multiple STTs together or supplement digital tracking with analog methods like journaling. This selective use can be seen as a strategy to mitigate the perceived risks and limitations of STTs.

Furthermore, survey 2 shows that GS participants engage in deeper reflection practices, often focusing on personal and contextual factors that affect their well-being. The lower trust in STTs leads GS participants to rely less on automated data and more on their personal insights and contextual understanding. This reliance on personal interpretation can foster a deeper engagement with their reflection practices, as they actively contemplate and analyze their data instead of passively accepting the technology’s outputs. For example, GS participants might critically assess how their mood, environment, and daily experiences influence the data recorded by STTs.

The thematic analysis of qualitative data from survey 2 highlights that GS participants’ reflection practices are more holistic and integrated with their cultural and social contexts. GS participants’ lower trust in STTs might drive them to consider a wider range of factors during reflection. They might incorporate cultural beliefs, social interactions, and emotional states into their reflective practices. This holistic approach contrasts with GN participants, who might focus more on the quantified metrics provided by STTs. For instance, a GS participant might reflect on how their community’s health practices influence their well-being, beyond just the numerical data from their fitness tracker. 

Illustrative quotes and examples further elucidate these points. It was mentioned by a GS participant that a WhatsApp chat was used to jot down daily notes and reflections, serving as a quick tool to complement the fitness app, which was not fully trusted to capture emotional states. Conversely, it was indicated by a GN participant that steps and sleep were tracked by a Fitbit, which was relied upon to assess well-being without much questioning of the data provided. Such examples underscore the differences in how GS and GN participants interact with STTs. For example, a digital calendar might be used by a GS participant to log activities and an analog diary to reflect on mood and interactions throughout the day. This combination allows for a critical evaluation of how emotional and social experiences impact physical activity, leading to more nuanced insights. In contrast, a GN participant might rely solely on their smartwatch to monitor daily steps and sleep patterns, using the app’s feedback to make adjustments without much additional personal reflection. In conclusion, the lower trust in STTs among GS participants leads to distinctive usage patterns and richer, more contextually grounded reflection practices. \textbf{We suggest that designers should recognize the diverse needs of users in STT design by acknowledging local health practices, user stress, and cultural values, especially in the Global South. }

\subsection{Quasi-Relationships with STTs} 

Our study reveals that contemporary STTs are far from neutral technological tools. Positioned between users and their environment, they shape and mediate how users perceive and interact with the world. While STTs can enhance certain interactions, they may also diminish other experiences, such as intuitive or felt sensations, that fall outside the norms of interaction design.

Most STTs are designed to provide users with data and features, like graphs, to interpret their daily performance. However, it is crucial to recognize that user experiences are influenced by diverse values, ethics, and moral concerns unique to different communities. Drawing on Ihde's framework of human-technology relations~\cite{GoodDesign,ihde2012technics}, we find that users often develop quasi-relationships with their STTs over time, akin to navigating a complex relationship with another person. Our findings highlight a conceptual and temporal mismatch for GS participants and their STTs similar to the findings of Bentvelzen et al.~\cite{TMRM}. First, GS participants seek more contextualized tracking experiences and prefer labor-intensive modalities like audio recordings and journaling over convenient data-driven smartphone applications. This preference for depth over convenience is evident in their choice of tracking mediums. Second, GS participants express frustration with the lack of flexibility in their STTs, particularly when notifications about pending goals feel generic and impersonal.

The two identified mismatches are better exemplified by a GS participant (P9-GS) who developed a complex relationship with their fitness goal-setting smartphone app over the course of a year. They experienced a quasi-love when achieving goals but felt stressed on days when the app failed to recognize contextual factors (e.g., their stress from a challenging day), leading to a quasi-hate relationship. The participants expressed a desire for a balanced approach, seeking both guidance and openness from their STTs. For instance, one participant preferred using Frankie for journaling and audio expression, while relying on omnitrack for guidance during overwhelming moments. They suggested enhancing Frankie with message playback for better self-reflection. In our perceived trust survey, we observed a spectrum of guidance and openness from multiple STTs, as participants exercised their desired agency by repurposing digital technologies to gain greater flexibility in tracking their daily experiences.

To address the challenge of balancing guidance and openness, designers should consider raising awareness of the conceptual-temporal mismatch inherent in STT usage. This could mitigate users' tendency to form quasi-relationships with their STTs. Our GS participants highlighted misalignment in STT notifications, but appreciated the overall fitness-tracking capability. The conceptual mismatch occurred when the STTs depicted daily life as predictable, while the temporal mismatch arose from inconvenient notification timing. To realign conceptual expectations, designers could facilitate a conceptual reset or support users in meaningfully abandoning STTs when necessary. For instance, allowing flexible tracking with Frankie and gradual integration of Omnitrack for goal pursuit could address this issue.\textbf{ We suggest that designers explore a symbiotic approach to tracking, combining devices that offer varied guidance levels and tracking flexibility, rather than adhering strictly to a goal-centric model.}

\subsection{Implications for Future Research}

Based on our findings and interpretations, we suggest two opportunities for future research.

\subsubsection{Reconsidering Quantified Self.}
In this study, our focus was specifically on two aspects within the broader self-tracking ecosystem: perceived trust and reflection practices. Despite our critical arguments for the reconsideration of QS as a globally dominant design norm for perceiving well-being, our methodology primarily relied on quantifying the opinions of our participants, supplemented by collecting some layers of qualitative data. In HCI research~\cite{Li2010,Rooksby2014}, the stages of using STTs—preparation, collection, integration, and action—comprehensively describe human-STT relationships. We propose that employing a more ethnographic approach could better capture these stages in people's local environments, resulting in richer data for understanding the plurality of designing and experiencing STTs. 

The age gap among our participants was notable, with GS participants generally younger, mirroring trends in the mean ages of their respective countries~\cite{hofstede2016culture}. This age discrepancy carries important implications, particularly concerning the use of universal STTs. We speculate that younger GS participants may be exposed to contemporary self-tracking practices influenced by experts before they have the opportunity to critically assess or unlearn them. \textbf{We suggest a need for designers to consider the impact of such practices on younger users, ensuring that STTs are accessible and beneficial without imposing expert-driven norms prematurely.}

\subsubsection{Supporting Home Environments.}
In the theme of reflection strategies and settings for self-improvement, we found that many participants preferred their homes as the ideal location to engage in reflection (see Figure~\ref{fig:Location_Reflection}). Extensive HCI research has explored interactive systems designed to improve well-being in home environments~\cite{Laina,Balance,LOOP,Rainmaker}, offering insight into questioning universal self-tracking norms. These systems promote openness, urging users to reflect on the alignment of data with their feelings rather than relying solely on authoritative sources. Examples that emphasize aesthetics and delayed gratification foster deeper reflection, striking a balance between action and reflection~\cite{Balance,IOBits}. Drawing on the insights from such systems can enhance well-being among GS participants in their local practices and homes. Designers can enrich data tracking expressivity by understanding the symbiotic relationship between humans and their homes, especially for GS participants. Principles of Soma Aesthetic Design~\cite{Soma} can assist in foregrounding local values and challenging data-centric approaches. As remote work blurs the boundaries between work and leisure at home, designers can create artifacts that foster mindfulness and delineate overlapping contexts.

\subsection{Limitations}

We recognize several limitations in our study. The relatively small sample sizes from each region constrain the generalizability of our findings. First, using countries as proxies for regions can oversimplify the diverse subcultures within each region, potentially providing diverse perspectives. Second, participants who have migrated between regions may have influenced our results. Lastly, our sample representing the global south consists of well-educated individuals fluent in English, potentially limiting the representativeness of their respective regions.

\section{Conclusion}

In response to the call for diversifying participant samples in HCI research, our study examines the influence of QS through STTs on individuals from diverse geographical backgrounds. We make a modest contribution to the broader theme of unequal power dynamics between STT users and developers, which can result in biased designs focusing on perceived trust and reflection practices. We conducted two iterative surveys with 156 and 121 participants. Our findings reveal significant disparities in perceived trust and reflection practices between the GS and GN participants. Advocating for inclusive and just design, we emphasize the importance of prioritizing local values and moving away from a one-size-fits-all approach. These results are consistent with previous research, indicating mismatches between user preferences and STT functionalities, particularly among GS participants. Future research opportunities include evaluating the impact of premature exposure to expert-driven STTs in younger GS individuals and using HCI insights to foster deeper reflection and engagement in diverse cultural contexts.

\appendix

\section{Survey 1: Perception of trust} \label{Premise}

\textbf{Premise}: Our daily lives can become very fast paced when trying to chase after goals that fulfill our desire of having a good life. While doing so, we build routines to help us be productive and efficient for the activities we do on a daily basis. Activities that we track can be very different for every person and can range from physical exercises to maintaining mindfulness. One may like to track these activities for reflecting and being aware of how they conduct their daily life. Smartphone applications and wearable technology like smartwatches have become popular in this regard. The following questions will be focused on this premise described and would help us understand how such technologies can be meaningful towards tracking for sefl-awareness.

\begin{figure}
    \centering
    \includegraphics[width=\linewidth]{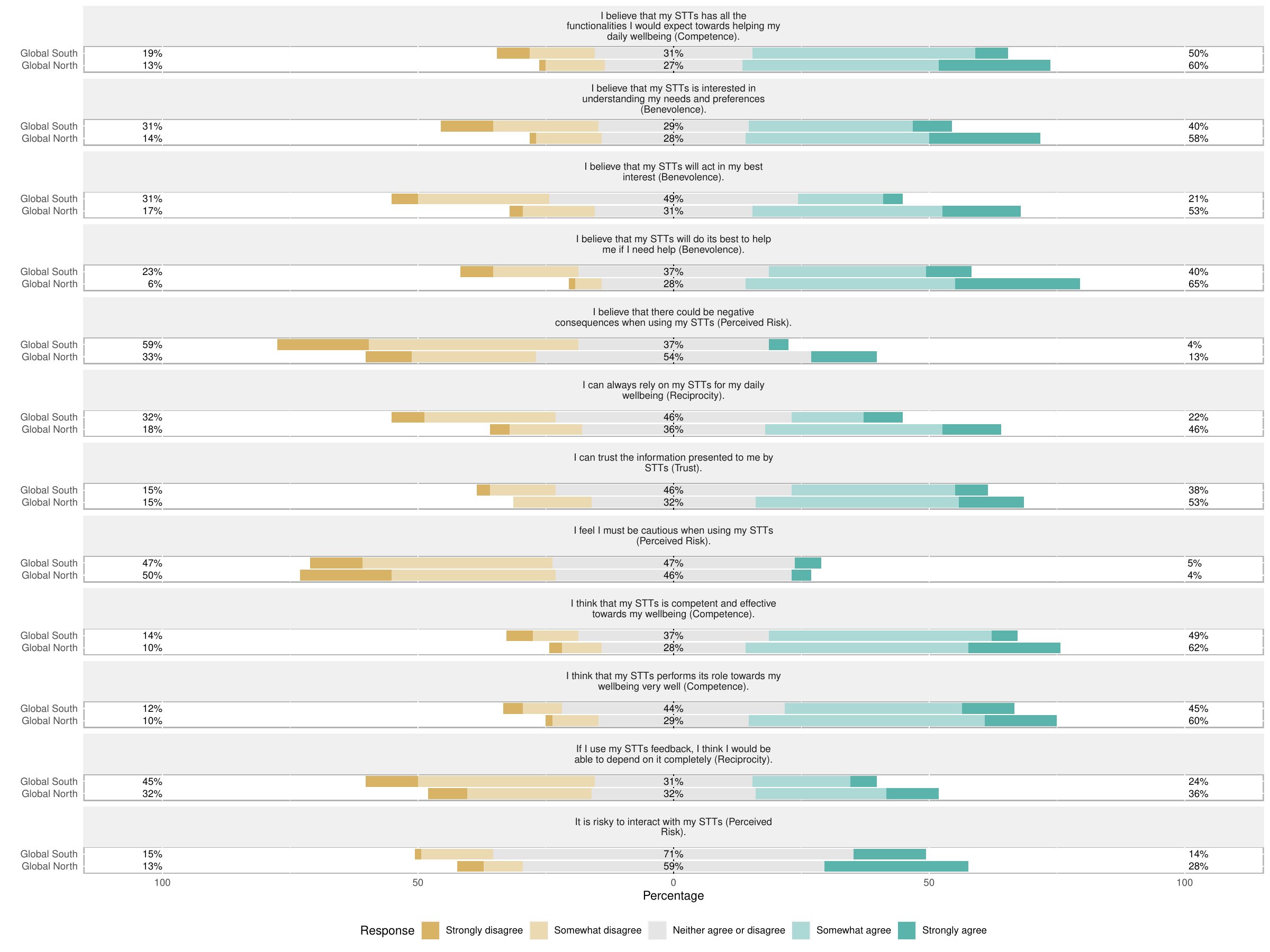}
    \caption{Percentage of responses to the individual Likert questions of the HCTS Survey between GS and GN-participants. The sub-scale are mentioned in brackets to which the statements belong.}
    \label{fig:HCTS}
    \Description{This image contains Likert scale responses to the 12 individual questions. The responses are shown in percentages as divergent bar graphs. Each Likert scale question shows the responses from both user groups.}
\end{figure}

\section{Survey 2: Reflection Practices} \label{Reflection}

\subsection{Reflection and well-being}    

     \begin{enumerate}
         \item What does well-being mean to you?
         \item What does reflection mean according to you?
         \item How often would you say do you reflect?
         \item How do you reflect?
         \item Does reflecting on your daily life support your well-being?
         \item Are you satisfied with this frequency?
         \item When do you reflect?
         \item Has there been a certain period in your life when you experienced a higher need for self-reflection?
         \item What are the causes for reflection? What triggers moments of reflection the most?
         \item Do you reflect individually or with others?
         \item Where do you reflect?
         \item What characterises your way of reflection?
         \item Please specify how reflection happens with your answers to the previous question.
     \end{enumerate}

\textbf{Technology and tools for reflection and activity tracking}

     \begin{enumerate}
         \item Do you use a tool or a technology (e.g. a journal, smart phone application, or other technology) to help you with your reflection?
         \item Please specify what is the well being tool and how you use it for your reflection moments?
         \item How many well-being tools do you currently use?
         \item How satisfied are you with your well-being tool(s) ?
         \item How long have you been using your well-being tool(s)?
         \item Have you faced any of these incidents (e.g. external validation, peer pressure, lack of autonomy, belittling, FitBit showing military locations) with your well-being tool?
         \item Could you suggest a product or system that can satisfy well-being by supporting reflective moments in everyday life? (It can be completely different from what you use).
     \end{enumerate}

\subsection{Qualities of technologies} 
Here people get to see 3 state of the art examples~\ref{fig:Survey_Picture}.

     \begin{enumerate}
         \item Which of these technologies to track data gives you the most flexibility?
         \item Please explain your decision.
         \item Which of these technologies offers the most freedom of input (e.g. how do you record / track the data, what do you record, etc.)?
         \item Please explain your decision.
         \item Which of these technologies offers you to track multiple types of data (e.g. not just how much or how often, but also why and how)?
         \item Please explain your decision.
     \end{enumerate}

\subsection{Open-Questions}
Having seen these technologies, please answer the following questions.

     \begin{enumerate}
         \item Which of the given example tools do you prefer for tracking well-being in your everyday life and why?
         \item How flexible is / are your well-being tool for recording your everyday activities? (eg. Choosing what you wish to record about the activity)
         \item Would you consider well-being tools to support or capture your emotions? Please explain your answer.
         \item Do you have suggestions for how your tool could capture emotions?
     \end{enumerate}

\end{document}